\documentclass{appolb}
\usepackage{graphicx}
\usepackage{graphicx}
\usepackage{dcolumn}
\usepackage{bm}
\usepackage{slashed}
\usepackage{amsmath,graphicx}
\usepackage[colorlinks=true,linktocpage=true,linkcolor=blue,citecolor=blue]{hyperref}
\usepackage{float}
\usepackage{nicefrac}
\usepackage[normalem]{ulem}
\usepackage{amsmath}
\usepackage{subfigure} 
\usepackage{bbold} 
\usepackage{placeins}
\usepackage[makeroom]{cancel}
\usepackage{stmaryrd} 

\def\be{\begin{equation}}
\def\ee{\end{equation}}
\newcommand{\bel}[1]{\begin{eqnarray}\label{#1}}
\newcommand{\eel}{\end{eqnarray}}
\def\barr{\begin{array}}
	\def\earr{\end{array}}
\def\beq{\begin{eqnarray}}
\def\eeq{\end{eqnarray}}
\def\bfig{\begin{figure}}
	\def\efig{\end{figure}}

\begin{document}
\title{Longitudinal spin polarization of $\Lambda$ hyperons in a thermal model%
\thanks{Presented at XXIXth International Conference on Ultra-relativistic Nucleus-Nucleus Collisions (QM 2022), April 4--10, 2022, Krakow, Poland}%
}
\author{Wojciech Florkowski
\address{Institute of Theoretical Physics, Jagiellonian University, 30-348 Krak\'ow, Poland}
\\[3mm]
{Avdhesh Kumar 
\address{Institute of Physics, Academia Sinica, Taipei, 11529, Taiwan}
}
\\[3mm]
Aleksas Mazeliauskas
\address{Theoretical Physics Department, CERN, 1211 Geneva 23, Switzerland}
\\[3mm]
Radoslaw Ryblewski
\address{Institute of Nuclear Physics, Polish Academy of Sciences, 31-342 Krak\'ow, Poland}
}
\maketitle
\begin{abstract}

We briefly review the thermal model predictions related to the longitudinal spin polarization of $\Lambda$ hyperons emitted from a hot and rotating hadronic medium produced in non-central relativistic heavy-ion collisions. 

\end{abstract}

\section{Introduction}

 Recent experiments by the STAR collaboration at RHIC, BNL have successfully measured the global polarization of the $\Lambda$, $\Xi$, and $\Omega$ particles~\cite{STAR:2018gyt,STAR:2020xbm}. Theoretically, the measured global spin polarization of the $\Lambda$ hyperons, integrated over all momenta, has been successfully described by the relativistic hydrodynamic models~\cite{Becattini:2016gvu,Karpenko:2016jyx}. In this approach, the basic quantity that dictates the spin polarization effects is the thermal vorticity tensor~\cite{Becattini:2013fla}. Surprisingly, predictions of hydrodynamic models~\cite{Becattini:2017gcx} based on the thermal vorticity fail to reproduce the local spin polarization (the momentum dependence of the longitudinal spin polarization) of $\Lambda$'s ~\cite{STAR:2019erd,Niida:2018hfw}. 

Recently, it has been suggested that thermal vorticity is not the only hydrodynamic quantity that contributes to the local spin polarization. In fact, symmetric hydrodynamic gradients defined by the quantity called the thermal shear, $\xi_{\mu\nu}=\frac{1}{2} (\partial_\mu \beta_\nu+\partial_\nu \beta_\mu)$, also contribute to the local spin polarization~\cite{Becattini:2021suc,Liu:2021uhn}.  Interestingly, by taking into account the thermal shear contribution,  hydrodynamic model predictions are found to be in agreement with the data only if the contribution from the temperature gradients in thermal vorticity and thermal shear are neglected~\cite{Becattini:2021iol} or the mass of the $\Lambda$ hyperon is replaced by the constituent strange quark mass~\cite{Fu:2021pok}.
 This kind of behavior motivated us to include the effect of thermal shear in our previous work~\cite{Florkowski:2019voj} that used thermal model with single freeze-out~\cite{Broniowski:2001we}.  In this contribution, we report our recent predictions on the longitudinal spin polarization arising from various contributions included in thermal vorticity and thermal shear within the thermal model with single freeze-out~\cite{Florkowski:2021xvy}.

\section{Thermal model, thermal vorticity and thermal shear tensors}
\label{sec:thm} 
A typical thermal model with single freeze-out uses four parameters: temperature $T$, baryon chemical potential $\mu_B$, proper time $\tau_f$, and system size $r_{\rm max}$. The thermodynamic parameters $T$, $\mu_B$ are fitted from the ratios of hadronic abundances measured in experiments. The geometric ones, $\tau_f$ and $r_{\rm max}$, characterize the freeze-out hypersurface (defined through the conditions: $\tau^2_f = t^2 - x^2 - y^2 - z^2$ and $x^2 + y^2 \leq r^2_{\rm max}$) and are obtained by the fits of experimental transverse-momentum spectra. The hydrodynamic flow is assumed to be of the Hubble-like form $u^\mu = x^\mu/\tau$. 
Herein, we use the extended thermal model with single freeze-out where the phenomena such as elliptic flow are included by taking into account the elliptic deformations of both the emission region in the transverse plane and of the transverse flow~\cite{Broniowski:2002wp} by means of the following parameterization 
\beq
x &=& r_{\rm max} \sqrt{1-\epsilon} \cos\phi, \quad
y = r_{\rm max} \sqrt{1+\epsilon} \sin\phi, \label{eq:xy}\\
u^\mu&=&\frac{1}{{\sqrt{\tau ^2-\left(x^2-y^2\right)\delta}}}\left(t,~x\sqrt{1+\delta},~ y\sqrt{1-\delta },~z\right), \label{eq:u}
\eeq
where $\phi$ is the azimuthal angle, while $\delta$ and $\epsilon$ are the model parameters. For $\epsilon >0$ the system formed in the collisions is elongated in the $y$ direction. The parameter $\delta$ accounts for the anisotropy in the transverse flow. For $\delta > 0$ there is more flow in the reaction plane (positive elliptic flow).  The parameter $\tau$ is the proper time, $\tau^2 = t^2 -x^2-y^2-z^2$. The values of the parameters $\epsilon$,   $\delta$, $\tau_f$, and $r_{\rm max}$ were determined in the past to describe the PHENIX data for three different centrality classes at the beam energy $\sqrt{s_{NN}}=130$~GeV for the freeze-out temperature $T_f =$~165~MeV~\cite{baran,Florkowski:2004du} and are listed in the table I of Ref.~\cite{Florkowski:2021xvy}. Equations (\ref{eq:xy}) and (\ref{eq:u}) can be directly used to obtain the velocity gradient terms  (${\varpi^{I} _{\mu \nu }}$, ${\xi^{I} _{\mu \nu }}$) of the thermal vorticity ($\varpi_{\mu\nu}$) and thermal shear ($\xi_{\mu\nu}$), as defined by Eqs. (7) and (11) of Ref.~\cite{Florkowski:2021xvy},  while the temperature gradient terms (${\varpi^{II} _{\mu \nu }}$, ${\xi^{II} _{\mu \nu }}$) 
 can be obtained using the expression $\partial^{\alpha }T=T \left(D u^{\alpha }- c_s^2 u^{\alpha }\partial_\mu u^{\mu }\right)$ (here $c_s=1/\sqrt{3}$ is the speed of sound) derived using the perfect-fluid hydrodynamic equations as discussed in the Appendix A of Ref~\cite{Florkowski:2019voj}.
\section{Predictions of various spin polarization observables}
The spin-polarization of particles with a given four-momentum $p$ is obtained by calculating the mean Pauli-Luba\'nski (PL) vector $\langle\pi^\star_{\mu}(p)\rangle$ in their local rest frame. The mean PL vector $\langle\pi_{\mu}(p)\rangle$ of particles with momentum $p$ emitted from a given freeze-out hypersurface is calculated by the formula
\begin{eqnarray}
\langle\pi_{\mu}(p)\rangle=\frac{E_p\frac{d\Pi _{\mu }(p)}{d^3 p}}{E_p\frac{d{\cal{N}}(p)}{d^3 p}}, \label{avPLV}
\end{eqnarray}
where 
\begin{eqnarray}
E_p\frac{d\Pi _{\mu }(p)}{d^3 p} &=& -\frac{ \cosh(\xi)}{(2 \pi )^3 m}
\int
e^{-\beta \cdot p} \,\Delta \Sigma \cdot p \,\,
\tilde{\omega }_{\mu \beta }p^{\beta }, \label{PDPLV}\\
E_p\frac{d{\cal{N}}(p)}{d^3 p}&=&
\frac{4 \cosh(\xi)}{(2 \pi )^3}
\int
e^{-\beta \cdot p} \,\Delta \Sigma \cdot p  
\,. \label{eq:MD}
\end{eqnarray}
Here $\Delta \Sigma _{\lambda}$ is an element of the freeze-out hypersurface. We assume that the spin polarization tensor $\omega^{\rho \sigma}$ at local equilibrium is given by the combination of thermal vorticity and thermal shear terms, ${\omega }^{\rho \sigma}=\varpi ^{\rho \sigma }+2 \hat{t}^{\rho }\frac{p_{\lambda }}{E_p}\xi^{\lambda \sigma }$ ~\cite{Becattini:2021suc,Becattini:2021iol}. Since we have already calculated various components of the thermal vorticity and thermal shear tensors in the thermal model, therefore,  we can determine various components of  the spin polarization tensor $\omega^{\rho \sigma}$. Finally, using the parameterization of the particle four-momentum $p^\lambda = \left(\sqrt{m^2 + p_T^2}\cosh y_p, p_x ,p_y, \sqrt{m^2 + p_T^2}\sinh y_p \right)$ in terms of rapidity $y_p$ and transverse momentum  $p_T = \sqrt{p_x^2+p_y^2}$, and assuming that the freeze-out takes place at constant value of the proper time ($\tau = \tau_f$) we can easily carry out integration over freeze-out hyper-surface and obtain $\langle\pi_{\mu}(p)\rangle$ defined by Eq.~(\ref{avPLV}). This formula can be subsequently used to plot  the $(p_x,p_y)$ dependence of the longitudinal spin polarization in the rest frame of particles ($\langle\pi^\star_{z}(p)\rangle$).

In recent experiments, the quantities of interest are:  the azimuthal-angle dependence of $p_T$-integrated longitudinal spin polarization $\langle P(\phi_p)\rangle$, $n=2$ azimuthal harmonic of the longitudinal spin polarization (polarization anisotropy) $\langle P_2\rangle$, and its transverse momentum dependence $\langle P_2(p_T)\rangle$. These quantities can be obtained using Eqs. (21), (20), and (22) of Ref.~\cite{Florkowski:2021xvy}, respectively. 
\section{\label{sec:results} Results}
 
\begin{figure*}
	\centering
	\subfigure[]{}\includegraphics[width=0.24\textwidth]{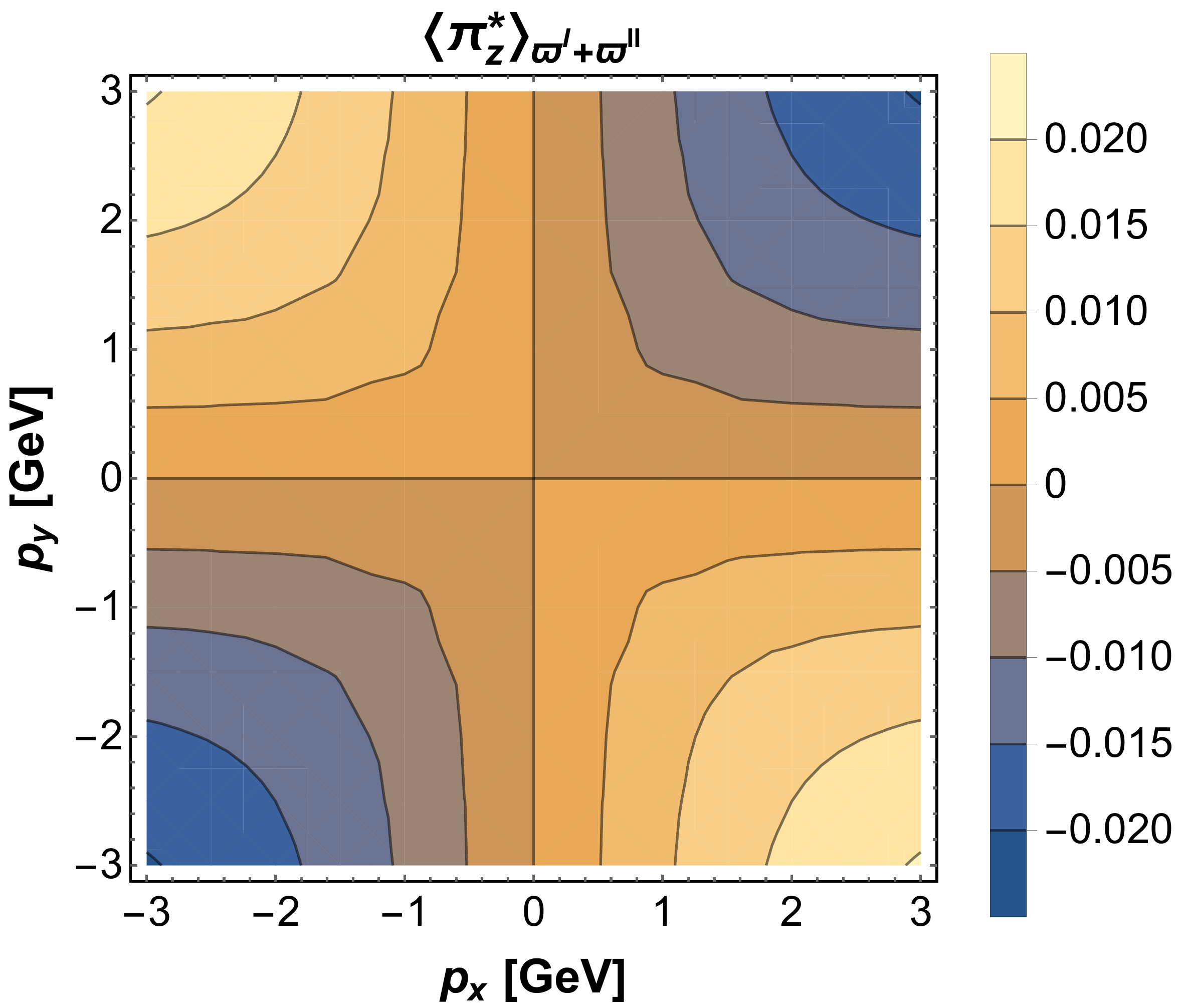}
	\subfigure[]{}\includegraphics[width=0.24\textwidth]{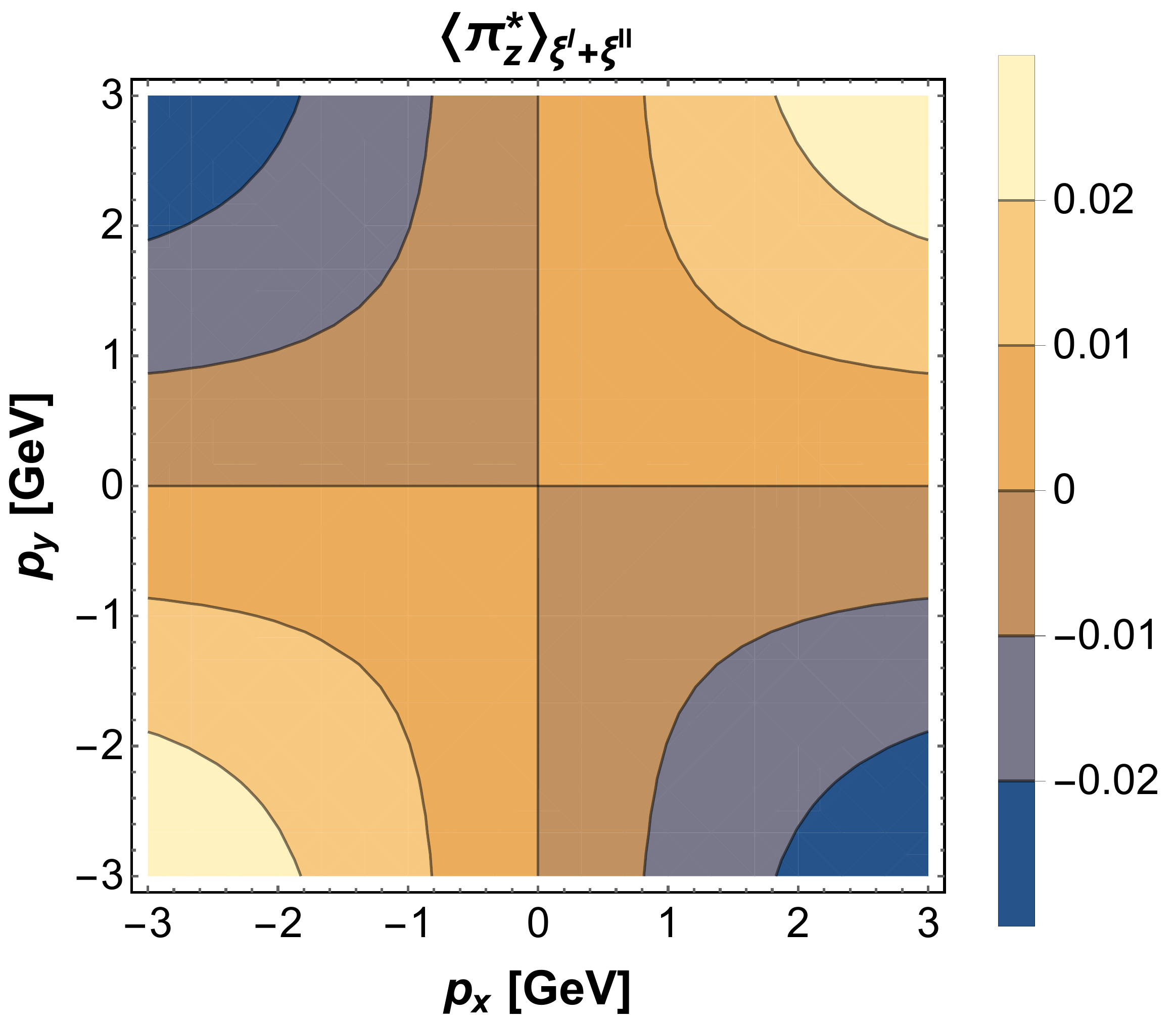}
	\subfigure[]{}\includegraphics[width=0.24\textwidth]{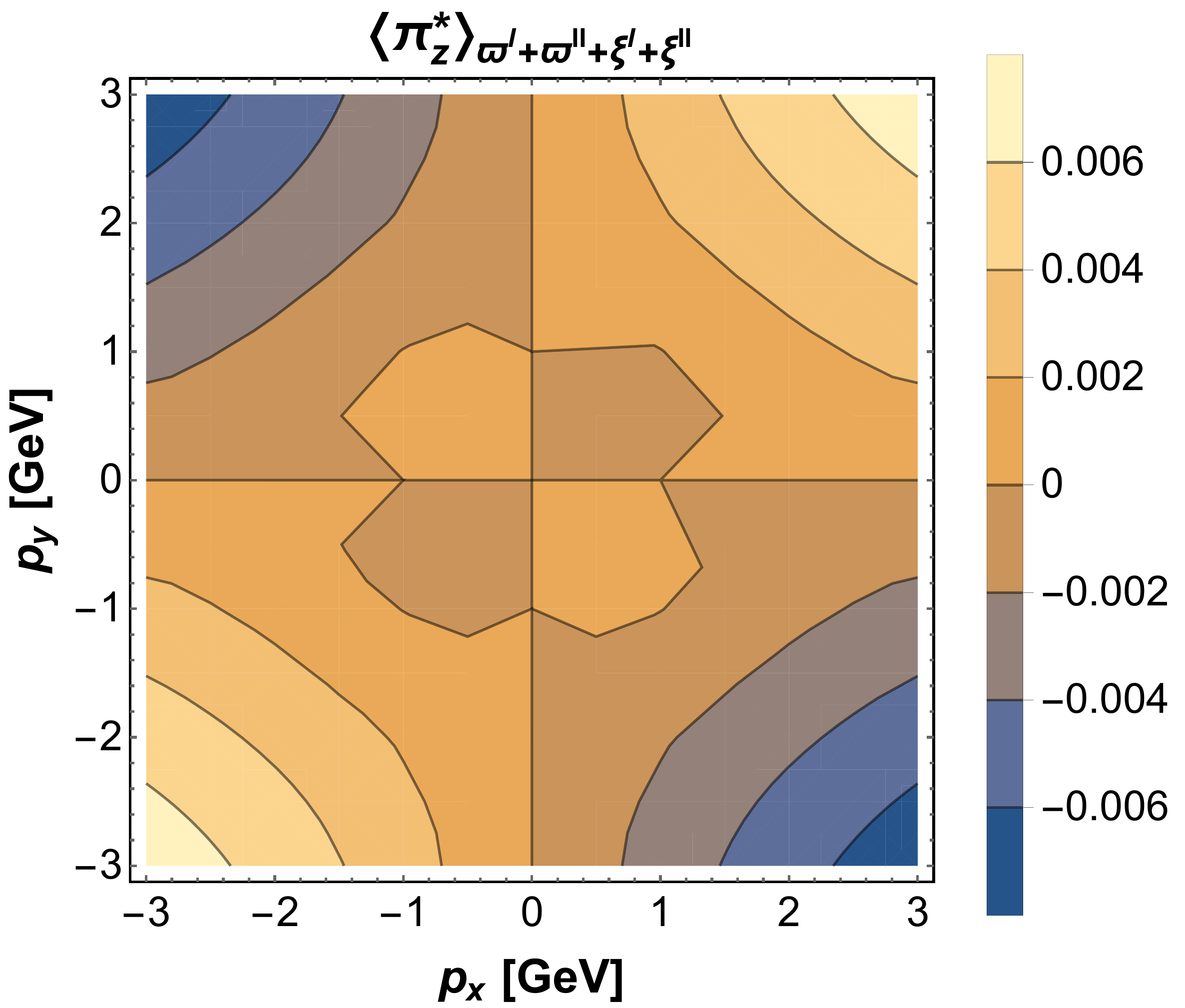}
	\subfigure[]{}\includegraphics[width=0.24\textwidth]{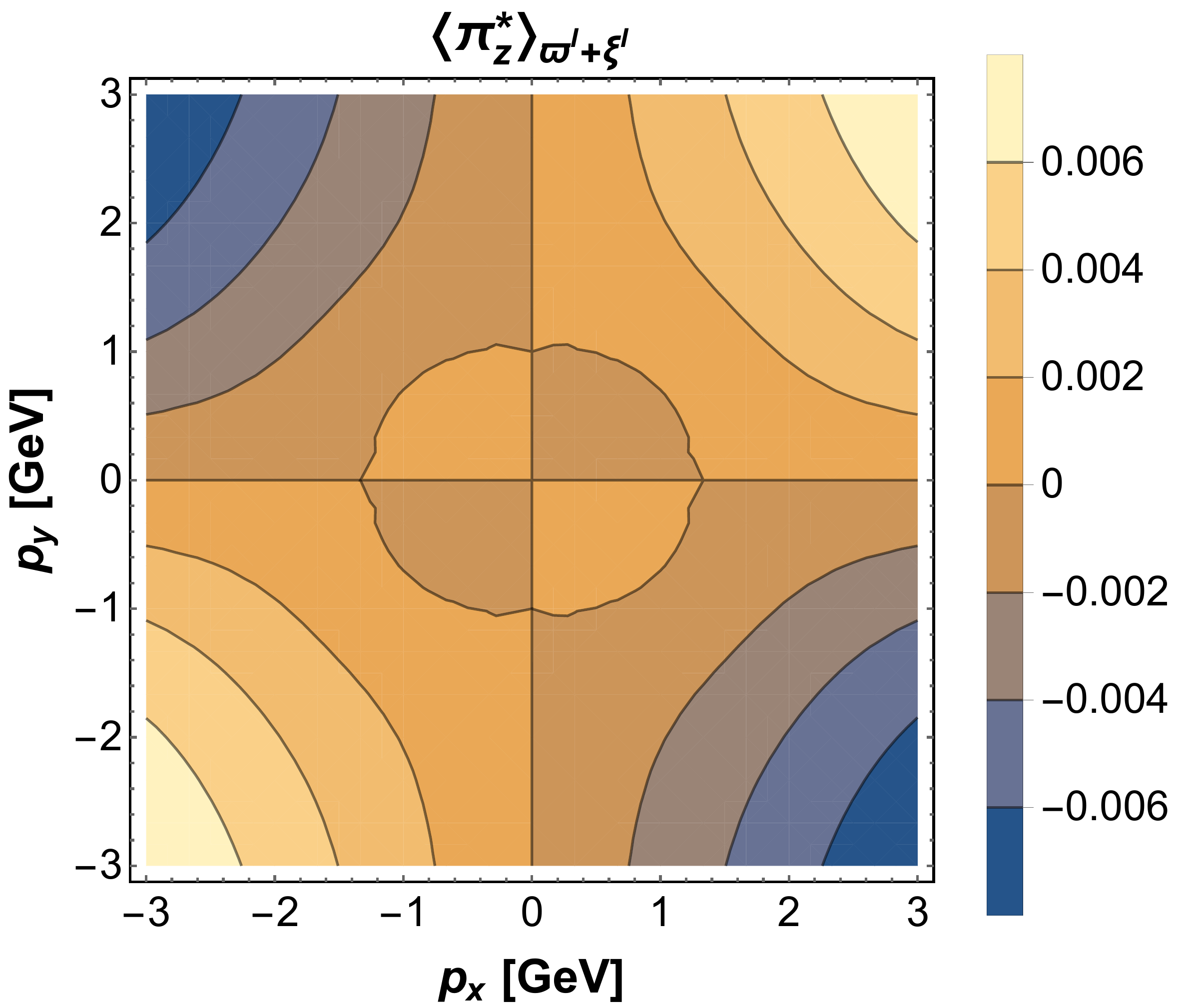}
	\caption{Contributions due to various terms in thermal vorticity and thermal shear to the longitudinal component of the mean spin polarization three-vector of $\Lambda$ hyperon as a function of its transverse momentum at $y_p=0$ with the model parameters used for the centrality class $c$=30--60\%  at $\sqrt{s_{NN}}=130$~GeV \,(see Table I of Ref.~\cite{Florkowski:2021xvy}).
	} 
	\label{fig:polarization1}
\end{figure*}

\begin{figure*}
	\centering
	\subfigure[]{}\includegraphics[width=0.46\textwidth]{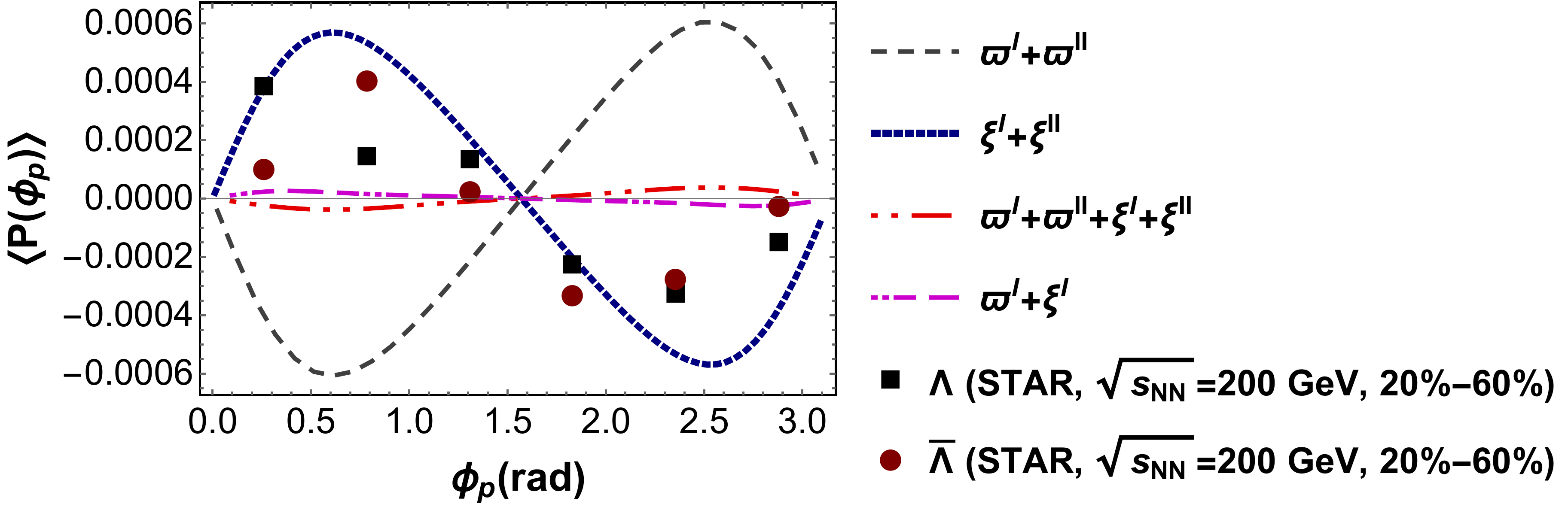}
	\subfigure[]{}\includegraphics[width=0.26\textwidth]{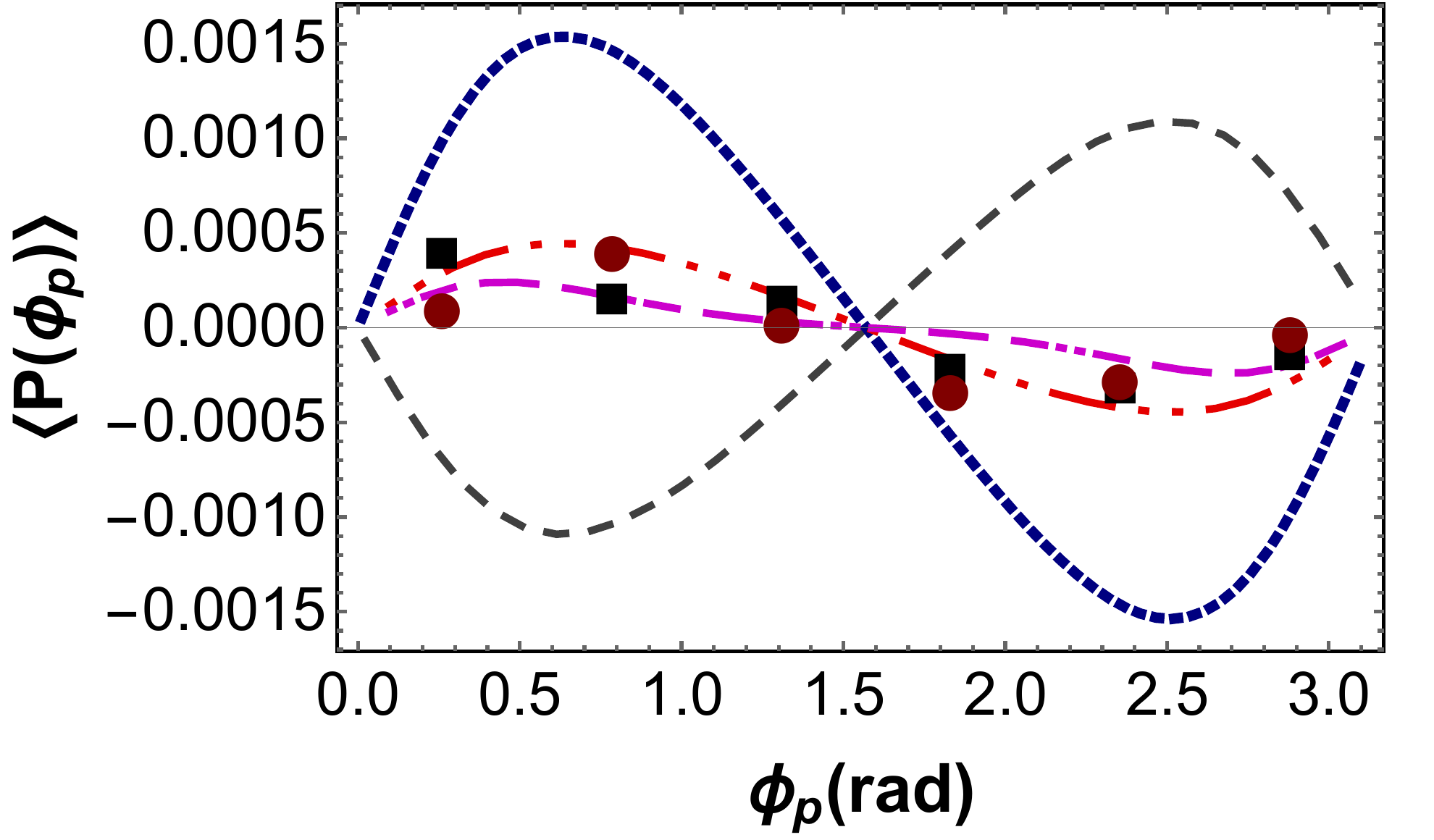}
	\subfigure[]{}\includegraphics[width=0.26\textwidth]{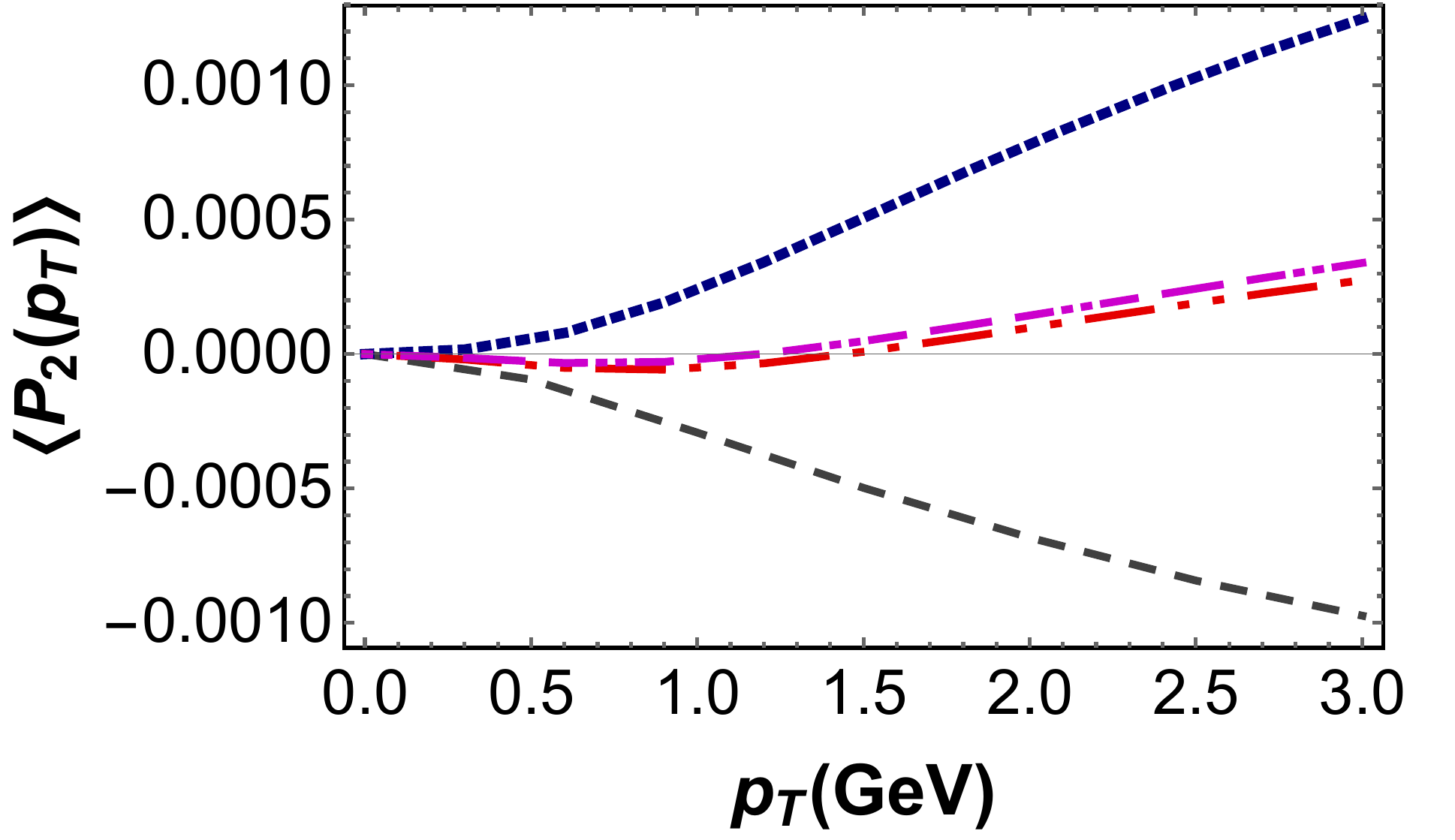}
	\caption{(a) and (b) show azimuthal angle dependence of $p_T$-integrated (range $p_T$=0--3 GeV) longitudinal spin polarization for $c$=30--60\% due to various terms of thermal vorticity and thermal shear tensor when the particle's mass is taken to be $\Lambda$--hyperons mass and s--quark mass.
		(c) Dependence of $n=2$ harmonic of longitudinal spin polarization on particle's transverse-momentum.
	}
	\label{fig:phipt}
\end{figure*}

Here we discuss our numerical results for the $p_x$, $p_y$ dependence of the longitudinal spin polarization ($\langle\pi^\star_{z}(p)\rangle$) of the $\Lambda$ hyperons (mass $m_{\Lambda}=$1.116 GeV), and other experimental observables as given by Eqs. (21), (20), and (22), respectively, in Ref.~\cite{Florkowski:2021xvy}  for the centrality class $c$=30--60\%  at $y_p=0$. From Figs.~\ref{fig:polarization1}(a) and \ref{fig:polarization1}(b)  we observe that  the thermal vorticity and thermal shear lead to opposite sign patterns of the quadrupole structure of the longitudinal component of the $\Lambda$ spin polarization, while their sum, as shown in  Fig.~\ref{fig:polarization1}(c), gives a sign pattern of the quadrupole structure such that it depends on the values $p_x$ and $p_y$. At larger  values of $p_x$ and $p_y$ we obtain the polarization sign pattern as observed in experiments. A similar pattern, as shown in Fig.~\ref{fig:polarization1}(d), is also observed for the net polarization when we neglect the temperature gradients.   
In Table~\ref{tab} we show our results for $\langle P_2 \rangle$ obtained by carrying out the $p_T$ integration in the range from 0 to 3 GeV. It can be noticed that the contributions from the total thermal vorticity and thermal shear to the $n=2$ harmonic of the longitudinal spin polarization are of opposite signs but close in magnitude. Their net contributions give a very small negative polarization value. If the temperature gradients are excluded, the sign of the net polarization changes, but the value remains an order of magnitude smaller than the individual contributions from the thermal vorticity or thermal shear. In Fig.~\ref{fig:phipt}(a) we present the plots for azimuthal-angle dependence of the $p_T$-integrated (range 0--3 GeV) longitudinal spin polarization $\langle P(\phi_p)\rangle$.  The black dashed and blue dotted curves correspond to the thermal vorticity and thermal shear contributions, while the red and purple dot-dashed curves show the net contribution with and without temperature gradients. It can be noticed that with temperature gradients included, the net result is practically zero. However, if the temperature gradients are excluded, we get the net polarization sign the same as from the thermal shear. To make a comparison with the experimental results, we also plot the experimental data for azimuthal angle dependence of the longitudinal spin polarization of $\Lambda$ and $\bar{\Lambda}$ for the centrality class $c$=20--60\% measured by STAR experiment at $\sqrt{s_\text{NN}}=200\,\text{GeV}$~\cite{STAR:2019erd}. The net results obtained from our model (with or without temperature gradients contribution) are too small to explain the experimental data. Next, following the idea discussed in Ref.~\cite{Fu:2021pok} we replace the mass of the Lambda hyperon ($m_{\Lambda}=1.116$ GeV) by the strange quark mass ($m_s=0.486$ GeV). In this case, as shown in Fig.~\ref{fig:phipt}(b), we are able to fit the data. In Fig.~\ref{fig:phipt} (c) we have shown the $p_T$-dependence of $n=2$ azimuthal harmonic of the longitudinal spin polarization $\langle P_2(p_T)\rangle$.   
As expected the contribution from thermal vorticity and thermal shear follow opposite trends. At low momenta ($p_T<1\,\text{GeV}$), the thermal vorticity contribution is larger while for larger momenta, $p_T>1\,\text{GeV}$, the thermal shear contribution becomes dominant.
\begin{table*}
	\centering
	\begin{tabular}{|p{1.6cm}||p{1.6cm}||p{1.8cm}||p{2.9cm}||p{1.8cm}|} 
		\hline
		c $\%$ & $\langle P_2\rangle_{\varpi^{I}+\varpi^{II}}$  &   $\langle P_2\rangle_{\xi^{I}+\xi^{II}}$  &  $\langle P_2\rangle_{\varpi^{I}+\varpi^{II}+\xi^{I}+\xi^{II}}$ & $\langle P_2\rangle_{\varpi^{I}+\xi^{I}}$ \\ 
		\hline 
		$30-60$ &   $-0.000292$ &  $0.000275$ & $-0.0000172$  &  $9.8\times10^{-6}$\\ 
		\hline
	\end{tabular}
	
	\caption{n=2 azimuthal harmonic due to various terms in thermal vorticity and thermal shear with model parameters used for $c$=30--60\% at $\sqrt{s_{NN}}=130$GeV~\cite{Florkowski:2021xvy}. 
		}
	\label{tab}
\end{table*}
\section{Summary and conclusions}
We briefly presented calculations of the various experimental observables related to the longitudinal spin polarization of $\Lambda$ hyperons in a single freeze-out thermal model. Our analysis suggest that the contributions of thermal vorticity and thermal shear to the azimuthal angle dependence of the longitudinal polarization are of opposite sign and nearly identically cancel each other leading to the disagreement with the data.   However, upon changing the mass of $\Lambda$--hyperons  to its constituent strange quark mass we found that our results agree with the experimental data. 

AK was supported in part by the Ministry of Science and	Technology, Taiwan, Grant No. MOST 110-2112-M-001-070-MY3. RR was supported in part by the Polish National Science Centre Grant No. 2018/30/E/ST2/00432.


\end{document}